\documentclass[authorcolumns]{lipics-v2021}
\pdfoutput=1
\usepackage{cite}
\usepackage{flushend}
\usepackage{todonotes}
\usepackage{algorithmic}
\usepackage{graphicx}
\usepackage{multirow}
\usepackage{xurl}
\usepackage[table,xcdraw]{xcolor}
\usepackage[breakable,many]{tcolorbox}
\usepackage{hyperref}
 \usepackage{todonotes}
\usepackage{balance}
 \usepackage{flushend}
\definecolor{main}{HTML}{CFCFCF}
\definecolor{sub}{HTML}{CFCFCF}
\tcbset{
    rounded corners,
    colback = white,
    before skip = 0.2cm,
    after skip = 0.5cm
}
\newtcolorbox{boxC}{
    colback = sub,
    boxrule = 0pt
}
\newcounter{keyTakeAwaysCounter}
\def\BibTeX{{\rm B\kern-.05em{\sc i\kern-.025em b}\kern-.08em
    T\kern-.1667em\lower.7ex\hbox{E}\kern-.125emX}}
\AtBeginDocument{
  \providecommand\BibTeX{{
    Bib\TeX}}}
\nolinenumbers
\begin{document}
\title{Identifying unique developers in OSS projects: A family of models}
\titlerunning{Identifying unique developers in OSS projects: A family of models}
\author{Ruoyu Su\footnote{Corresponding author}}{University of Oulu, Finland}{ruoyu.su@oulu.fi}{https://orcid.org/0009-0008-6206-8787}{}
\author{Alexander Bakhtin}{University of Oulu, Finland}{alexander.bakhtin@oulu.fi}{https://orcid.org/0000-0003-3513-7253}{}
\author{Matteo Esposito}{University of Oulu, Finland}{matteo.esposito@oulu.fi}{https://orcid.org/0000-0002-8451-3668}{}
\author{Davide Taibi}{University of Southern Denmark, Vejle, Denmark \and University of Oulu, Finland}{taibi@imada.sdu.dk}{https://orcid.org/0000-0002-3210-3990}{}
\author{Valentina Lenarduzzi}{University of Southern Denmark, Vejle, Denmark \and University of Oulu, Finland}{lenarduzzi@imada.sdu.dk}{https://orcid.org/0000-0003-0511-5133}{}
\authorrunning{R. Su, A. Bakhtin, M. Esposito, D. Taibi, and V. Lenarduzzi}
\Copyright{R. Su, A. Bakhtin, M. Esposito, D. Taibi, and V. Lenarduzzi}
\funding{
Research Council of Finland (grants no. 359861 and 349488 - MuFAno), Business Finland
(grant 6GSoft), and FAST, the Finnish Software Engineering Doctoral Research Network.}
\maketitle
\begin{abstract}
Organizational and logical coupling metrics require reliable identification of unique developers. In OSS, commit metadata is limited to names and emails, and the same developer may appear under multiple aliases, which can distort coupling measurements if de-duplication is missing.
We aim to build a scalable and accurate pipeline for OSS developer de-duplication and to provide guidance on choosing a model based on precision vs. computational effort.
We use Indel similarity as a baseline, then run an LLM-assisted matching process with manual validation to create a large dataset of duplicate identities. Using this dataset, we train and compare classical ML models of different complexity, evaluating precision along with training and inference time and energy.
We expect a high-quality dataset and a benchmark of approaches that clarifies which solutions offer the best trade-off between accuracy and cost for large-scale OSS mining.
\end{abstract}
\section{Introduction}

Recent research in software architecture has increasingly emphasized the socio-technical dimension of systems, highlighting how developer collaboration structures influence architectural properties such as coupling, cohesion, ownership, and system evolution. A growing body of empirical work has investigated organizational and logical coupling in service-oriented and microservice-based systems by mining developer collaboration networks and contribution histories~\cite{li2023analyzing, li2023evaluating, li2023metrics, li2024collaboration}. These studies demonstrate that architectural quality cannot be fully understood without accounting for how developers interact with architectural elements over time.

Subsequent investigations have explored architectural implications of developer allocation strategies, ownership patterns, and collaboration dynamics, showing how organizational factors may either amplify or mitigate architectural degradation~\cite{amoroso2023one, darioamoroso2023microservice, darioamoroso2024understanding, li2024framework, li2025toward, li2025exploring}. Across these works, \textbf{developer-centric metrics} are repeatedly used as \textbf{first-class inputs} for architectural analysis, optimization, and decision-making.

However, all such analyses fundamentally rely on the \textbf{assumption} that developer identities extracted from version control systems \textbf{accurately represent unique natural persons}. In open-source software repositories, this assumption rarely holds: the same developer may appear under multiple name–email combinations due to configuration changes, role-based accounts, or inconsistent identity usage. When left unresolved, duplicate developer identities introduce \textbf{systematic noise} into collaboration networks and coupling metrics, potentially distorting architectural interpretations and downstream empirical conclusions.
Despite its importance, developer identity de-duplication is often handled heuristically \cite{li2016task,das2024exploring}, performed manually on small datasets \cite{li2023analyzing,li2023evaluating,bakhtin2024temporal}, or omitted altogether \cite{amoroso2023one,li2024collaboration,li2025exploring,barcellini2008user}, implicitly treated as a \textbf{secondary data-cleaning concern}. As architectural research increasingly targets large-scale systems and longitudinal analyses, the lack of scalable, precise, and reproducible identity resolution mechanisms becomes a critical methodological limitation.

In this registered report, we aim to systematically investigate whether modern Large Language Models and classical Machine Learning approaches can provide accurate, scalable, and energy-aware solutions for duplicate developer identification in large-scale open-source projects, thereby strengthening the methodological foundations of developer-centric software architecture research. Henceforth, we expect this registered report to provide the following contributions:

\begin{itemize}
  \item A systematic empirical evaluation of Large Language Models for duplicate developer identity identification in open-source software repositories;
  \item A comparative assessment of classical ML-based approaches trained on LLM-curated data, focusing on precision, scalability, and computational efficiency;
  \item An analysis of time and energy consumption trade-offs between LLM-based and ML-based solutions in large-scale mining scenarios;
  \item An empirical investigation of how training data volume affects the precision of ML-based identity resolution approaches;
  \item An assessment of whether developer activity similarity (via cosine similarity) improves duplicate identity detection;
  \item A reusable and openly available experimental protocol and replication package to support future architectural and socio-technical mining studies.
\end{itemize}
\section{Background and Related Work}
\label{sec:back}

Mining commit data provides us with the name and email of the author/committer as they were set in that user's git configuration. This creates an issue of duplicate developer identities in the data, since the same person might forget to properly configure this data on different machines or simply change emails.
For instance, the organizational coupling metric by Li et al. \cite{li2023evaluating} involves counting contribution switches done between a pair of components by each developer, and weighing this count by the harmonic mean of churn incurred to both components by the same developer. Such a non-linear metric will yield different values when two identities, A and B, and their churns are used separately, compared to when all the churn incurred by A and B is attributed to the same merged identity.
Thus, given the full list of developer names and emails from mined commit data, we are required to perform developer de-duplication.

Many existing works constructing developer social networks through mining software repositories do not deal with this issue and do not address it as a threat to the validity of the work \cite{amoroso2023one,li2024collaboration,li2025exploring}, apparently treating each unique record as a unique developer, with some works acknowledging it openly \cite{barcellini2008user}. Some authors leveraged proprietary industrial data, where arguably more strict rules regarding configurations and committing to projects were in place and identity of each developer can be reliably established \cite{meneely2008predicting,wolf2009predicting,jermakovics2011mining,zhao2024openrank}, while others gathered data from official database dumps, which include a \emph{users} table acting as a foreign key for other records involving users \cite{li2016task,das2024exploring}. Finally, some works only gathered a small set of developers, so identification of inconsistencies and merging of identities were performed manually \cite{li2023analyzing,li2023evaluating, bakhtin2024temporal,li2025toward}.
The heuristic by Bird et al. \cite{bird2006mining} is the most commonly cited method for identifying unique $(name, email)$ pairs, although it was proposed for email lists \cite{ngamkajornwiwat2008exploratory,wang2021characterizing}.
It leverages Levenshtein similarity \cite{Levenshtein1966}, which is a string similarity metric that counts the number of additions, deletions, and substitutions necessary to transform one string into the other.

To our knowledge, this is the first attempt to apply Large Language Models to the problem of unique developer identification in OSS projects.

\section{Empirical study design}
\label{sec:method}

The empirical study design follows the guidelines defined by \cite{wohlin2012experimentation}. In this section, we describe the goals and research questions with hypotheses, the study context, the execution plan, and the data analysis.
Due to space constraints in the registered report, we uploaded the extended methodology, which includes a Figure describing the entire process and introducing LLM integration, design, and validation in the online appendix\footnote{\url{https://doi.org/10.5281/zenodo.19472235}}.

\subsection{Goal, Research Questions, and Hypothesis}

The \textbf{goal} of this empirical study is to \textit{construct} a family of ML models for duplicate developer identification \textit{for the purpose of} developer identity de-duplication \textit{with respect to} the precision and effort of the model inference, \textit{from the point of view of} researchers \textit{in the context of} mining OSS repositories.

In particular, we first investigate whether LLMs can reliably identify duplicate developer identities and generate labeled identity pairs. Then we use labeled pairs to support the training and evaluation of ML-based approaches for scalable duplicate developer identification.
Therefore, we define the following \textbf{Research Questions} (\textit{RQs}):

\begin{boxC}
\textbf{RQ$_1$}:
Can LLMs identify duplicate developer identities in OSS projects?
\end{boxC}

Identifying unique developer identities is a prerequisite for accurately computing organizational and logical coupling in OSS projects \cite{li2025exploring}; however, the developer identity de-duplication is complex and challenging due to the lack of authoritative identity information \cite{fry2020dataset}: inconsistent names and emails in the submission record; and traditional string-based heuristics rely on superficial similarity and can not identify the semantic variations, such as nickname, email format, and naming convention. Recent advances in LLMs suggest they may be able to resolve ambiguous developer identities by leveraging contextual and semantic information beyond traditional string-based heuristics \cite{li2024leveraging,huang2024leveraging}.
This \textit{RQ} aims to investigate whether LLMs can identify duplicate developer identities. Incorrectly identifying two different developers as the same person would undermine developer-based analysis, such as organizational coupling and collaboration network. In this study, LLMs serve as the decision-maker that receives the developer identity pair $(name, email)$ to determine whether they are the same natural person.

Considering the increasing popularity of diverse LLMs, we plan to leverage multiple LLMs to complete the duplicate developer identification task and examine whether the outcomes show statistically significant differences. Therefore, we conjecture the following \textbf{null} hypothesis (\textit{H$_0$}) and \textbf{alternative} hypothesis (\textit{H$_1$}):

\begin{itemize}
    \item \textit{H$_{01}$: There is no significant difference between LLMs in duplicate developer identification.}
    \item \textit{H$_{11}$: There is a significant difference between  LLMs in duplicate developer identification.}
\end{itemize}

\begin{boxC}
\textbf{RQ$_2$}: Can ML-based approaches provide a practical and scalable solution for duplicate developer identification in OSS projects?

\begin{itemize}
    \item \textbf{RQ$_{2.1}$}: Can ML-based approaches identify duplicate developer identities?
    \item \textbf{RQ$_{2.2}$}: Can ML-based approaches outperform LLMs in terms of time and energy consumption?

    \item \textbf{RQ$_{2.3}$}: Does the precision of ML-based approaches depend on the amount of training data?

\end{itemize}
\end{boxC}

While LLMs may identify duplicate developer identities (\textit{RQ$_1$}), their computational cost, limited scalability, and energy consumption may hinder their direct adoption in large-scale mining studies \cite{strubell2019energy,fernandez2025energy}. In practice, OSS repositories often contain thousands of contributors and millions of commits, with vast amounts of data \cite{MaesBermejo2024LinuxKernel}. \textit{RQ$_2$} aims to investigate whether classical ML-based approaches trained on LLM-curated data can provide a practical and scalable solution for duplicate developer identification. \textit{RQ$_2$} mainly focuses on evaluating whether ML-based approaches can effectively identify developers while being more efficient and easier to deploy at scale.

\textit{RQ$_{2.1}$} aims to determine the effectiveness of classical ML-based approaches in identifying duplicate developer identities. Different from LLMs, ML-based approaches rely on supervised learning with labeled identity pairs \cite{christen2012data}.
Therefore, their performance depends on the quality of the labeled dataset of developer duplicate pairs, which we aim to obtain as the result of \textit{RQ$_{1}$}.
Thus, in this sub-question, we aim to see which ML models pre-trained on the LLM-provided labels are good enough for developer de-duplication. Availability of pre-trained models as a result of this RQ will remove the burden of data curation and model training for those wishing to leverage our approach in the future, requiring time and resources only for the inference of the provided new data.

\textit{RQ$_{2.2}$} investigates the performance of ML-based approaches in terms of time and energy consumption. Even if ML-based approaches can identify developer identities, their practical applicability depends on their computational and energy efficiency \cite{tschand2025mlperf}. Therefore, facing massive amounts of data, training and inference costs become important factors in determining whether a method is practically deployable. \textit{RQ$_{2.2}$} evaluates the time and energy consumption of different ML-based approaches through quantifying these factors, compares them to the LLM-based plan (\textit{RQ$_1$}), and determines their suitability for large-scale empirical mining studies.
\textit{RQ$_{2.3}$} examines the relationship between training data volume and the precision of using ML-based approaches for duplicate developer identification. The availability of labeled data is another key limitation in developer identity resolution \cite{amreen2020alfaa}. Creating large, manually validated datasets is expensive and time-consuming, and it is unclear how much training data is required to achieve stable and reliable performance. \textit{RQ$_{2.3}$} explores whether the precision of ML-based approaches is associated with the the amount of training data: increasing, remaining unchanged, or decreasing.

So, we conjecture the following \textbf{null} hypotheses (\textit{H$_0$}) and \textbf{alternative} hypotheses (\textit{H$_1$}):

\begin{itemize}
    \item \textit{H$_{02}$: There is no significant difference between ML-based approaches in identifying duplicate developer identities.}

    \item \textit{H$_{03}$: There is no significant difference in time and energy consumption between ML-based approaches and LLMs for duplicate developer identification.}

    \item \textit{H$_{04}$: There is no significant association between the amount of training data and the precision of ML-based duplicate developer identification.}

    \item \textit{H$_{12}$: There is a significant difference between ML-based approaches in identifying duplicate developer identities.}

    \item \textit{H$_{13}$: There is a significant difference in time and energy consumption between ML-based approaches and LLMs for duplicate developer identification.}

    \item \textit{H$_{14}$: There is a significant association between the amount of training data and the precision of ML-based duplicate developer identification.}

\end{itemize}

\begin{boxC}
\textbf{RQ$_3$}:
Can cosine similarity between developers improve the quality of duplicate developer identification?

\end{boxC}

While the basic data we focus on in this work is the names and emails of developers, mining the commit history allows us to obtain information about the file modifications they made.
Jermakovics et al. \cite{jermakovics2011mining} proposed constructing a matrix indicating which developers committed to which files and using cosine similarity between developers' activity vectors to cluster them based on similar activity.
Cosine similarity refers to the similarity between developers’ activity profiles derived from their file modification histories.
We suspect that the same person, committing under several identities, would nonetheless touch related files, and thus the activity patterns of these identities would be very similar.
Thus, we can leverage this cosine similarity as an additional feature to provide to the considered methods for identifying duplicates and evaluate whether providing this feature results in a significant improvement in the methods' quality.
\textit{RQ$_3$} investigates whether adding cosine similarity information can improve the quality of identifying duplicate developer identities. This question is cross-cutting and applies to both LLM-based (\textit{RQ$_1$}) and ML-based approaches (\textit{RQ$_2$}). Through the comparison, we can evaluate whether adding this complexity is reasonable in large-scale empirical mining studies. So, we conjecture the following \textbf{null} (\textit{H$_0$}) and \textbf{alternative} hypothesis (\textit{H$_1$}):

\begin{itemize}
    \item \textit{H$_{05}$: There is no significant difference in the quality of duplicate developer identification when cosine similarity between developers is incorporated, compared to when it is not.}

    \item \textit{H$_{15}$: Incorporating cosine similarity between developers has a significant difference in the quality of duplicate developer identification.}
\end{itemize}

\subsection{Study Context}
As context, we plan to use the Linux kernel\footnote{\url{https://github.com/torvalds/linux}} as the data source of developer identities for our study. According to the project's GitHub mirror, the Linux kernel is one of the largest OSS software projects, with over a million commits and 15,000 contributors. Developers worldwide contribute to this project, ensuring the dataset of developer names and email addresses includes representative names from diverse cultural backgrounds and naming conventions. This enables us to evaluate the effectiveness of developer deduplication solutions across a diverse dataset.
While the Linux kernel project is the primary study context, our evaluation focuses on the reliable precision estimates. If the Linux kernel data does not provide sufficient duplicate identity cases to support robust precision analysis or model training, we will also mine other large-scale OSS projects with diverse contributor bases.

\subsection{Study Execution}

We describe our study execution plan, including the data collection and analysis strategies.

\textit{1) Linux Kernel Data Collection:} We clone the Linux kernel repository via GitHub mirrors. Each commit in Git version history contains a committer identity and an author identity, and we extract the developer identity from both. We represent each developer identity as a $(name, email)$ format in the commit metadata. We record the total number of commits and unique identities included in the study. After extraction, we obtain a set of $(name, email)$ identities.

\textit{2) Developer Identity Data Preprocessing:} After collecting developer identity information, we perform data preprocessing steps to clean and standardize the extracted identity data.
This step aims to remove minor inconsistencies and non-informative entities.
Specifically, we standardize developer names and email addresses by cutting leading and ending spaces.
We exclude developer identities with missing or empty email fields; we retain entries with missing or empty name fields, as emails may still contain identifiable naming information.
This stage does not involve any semantic transformation operations (extension or merging).

\textit{3) Identity Pair Generation:}
We generate a complete set of all pairs of mined identities for the subsequent duplicate developer identification.

Considering the number of unordered identity pairs grows quadratically with the number of developer identities, it may be computationally infeasible to evaluate all possible combinations in large OSS repositories.
Therefore, if the computational load is excessive, we apply a blocking strategy to determine the set of candidate pairs for which LLM processing is required, using low-cost string-based heuristics, such as Levenshtein similarity \cite{Levenshtein1966}, for names and emails.
For instance, the names of present authors “Ruoyu Su” and “Davide Taibi” do not share any letters of the alphabet, and thus their similarity would be close to 0. Thus, such pairs can be filtered before LLM processing if necessary, significantly decreasing the number of pairs passed to LLMs.
The final filtered identity pairs after blocking will be used for all the subsequent analyses.

\textit{4) Computational Environment and Execution Setup:} We run all experiments on the supercomputer hosted by CSC\footnote{\url{https://csc.fi/}}, the Finnish IT Center for Science. Roihu\footnote{\url{https://docs.csc.fi/computing/systems-roihu/}} is a high-performance computing system from CSC designed for compute- and data-intensive research with a total of 486 CPU nodes and 132 GPU nodes.
It has a heterogeneous architecture that each GPU node has 4 NVIDIA GH200 Grace Hopper superchips, and each GH200 superchip comprises one Hopper (H100) GPU and one Grace CPU with 72 ARM CPU cores. Roihu can have efficient and scalable execution for demanding AI and data-intensive applications.

\textit{5) LLM-Based Duplicate Developer Identification (RQ$_1$):} We employ multiple LLMs as independent decision-makers for duplicate developer identification. Each LLM receives the same input, the developer identity pair $(name, email)$, and determines whether these two identities belong to the same natural person.
We use the same prompt and consistent LLM parameters when performing duplicate developer identification with different LLMs to ensure fairness in cross-model comparisons.
Each LLM generates a binary decision for each candidate identity pair. We collect and compare outputs from different LLMs to test whether there is a significant difference in their duplicate developer identification decisions. To evaluate identification quality, human experts assess the correctness of selected LLM outputs. Human judgments are only for evaluation purposes (e.g., precision) and do not influence LLM outputs and comparisons between LLMs. To mitigate the bias introduced by evaluators' own experiences and uncertainty, such as different developers sharing similar names or emails, evaluators will be instructed to follow a conservative approach, where any uncertainty in duplicate detection should result in a pair being labeled as non-duplicate.

\textit{6) ML-based Approaches Duplicate Developer identification (RQ$_2$):} We evaluate practical and scalable solutions that leverage ML-based approaches for duplicate developer identification. We train ML models via supervised learning, using curated and labeled identity pairs from the LLMs' outputs in \textit{RQ$_1$}. ML-based approaches rely on the structural characteristics of developers' names and emails and are applied to the same identity pairs as in \textit{RQ$_1$} to maintain consistency across experiments. We consider multiple classical ML-based approaches, including record linkage libraries (e.g., Dedupe), string similarity-based approaches, and string embedding-based approaches. For each approach, we follow the same procedure for training and application independently.
To examine the effectiveness of ML-based approaches \textit{(RQ$_{2.1}$)}, we train ML models using the LLM-curated labeled identity pairs and perform duplicate developer identification. To evaluate time and energy consumption (performance) \textit{(RQ$_{2.2}$)}, we record the training and computational costs, including execution time and resource utilization, under the same computational environment described above, following CSC guidelines for GPU utilization monitoring\footnote{\url{https://docs.csc.fi/support/tutorials/gpu-ml/\#gpu-utilization}}. In addition, to investigate the dependence of ML-based approaches on training data \textit{(RQ$_{2.3}$)}, we set different amounts of training data to run the same training procedure and explore whether the scale of the training data affects identification precision.

\textit{7) Cosine Similarity Augmentation (RQ$_3$):} We investigate whether adding cosine similarity based on developers' activity profiles could improve the precision of duplicate developer identification. We extract the developers' activity profiles and calculate the cosine similarity between developer activity pairs. We integrate cosine similarity as an additional feature into both LLM-based (\textit{RQ$_1$}) and ML-based (\textit{RQ$_2$}) duplicate developer identification experiments. For LLM-based identification, we add cosine similarity as a contextual input alongside the $(name, email)$ identity pair. For ML-based identification, we include cosine similarity as an additional input feature during training and inference. We keep all conditions consistent except for adding cosine similarity to ensure a controlled comparison in the experiment. Consequently, we can evaluate whether adding cosine similarity information improves precision without additional confounding factors.

\subsection{Data Analysis}

Since this is a registered report, we did not execute the study and cannot know a \textit{priori} whether the collected data follow the normal distribution. Consequently, we define the data analysis protocol for both normal and non-normal cases in this section.

\textit{1) LLM-Based Duplicate Developer Identification (RQ$_1$):} We analyze the results of duplicate developer identification between LLMs. We consider \textit{precision} as the primary evaluation metric because it directly reflects the false-positive error rate. We also measure other standard metrics like \textit{recall} and \textit{F1 score} as complementary indicators.

To evaluate the correctness of LLMs for duplicate developer identification, we employ Cochran's formula \cite{israel1992determining} to obtain a sample size of developer identity pairs with a 95\% confidence interval and a 5\% error margin. Human experts independently label them as the evaluation benchmark. For each identity pair, human experts label whether the two developer identities are duplicates.

For each LLM, \textit{the precision} calculation formula is: $\mathrm{P} = \frac{TP}{TP + FP}$. \textit{TP} represents the number of correctly identified duplicate identity pairs, while \textit{FP} represents the number of incorrectly identified duplicate identity pairs. We also compute \textit{recall} and \textit{F1 score} based on the same labeled identity pairs.
All LLMs evaluate the same identity pairs and generate paired binary decisions. Therefore, we use McNemar's paired statistical test~\cite{dietterich1998approximate} to assess whether there is a significant difference between LLMs in duplicate developer identification.
We conduct the hypothesis testing at a significance level (e.g., $\alpha = 0.05$).

The null hypothesis \textit{(H$_{01}$)} states that there is no significant difference between LLMs in duplicate developer identification, while the alternative hypothesis \textit{(H$_{11}$)} states that at least one LLM differs significantly from others.

\textit{2) ML-based Approaches Duplicate Developer identification (RQ$_2$):} We analyze the results of duplicate developer identification between different ML-based approaches (\textit{RQ$_{2.1}$}). Consistent with \textit{RQ$_1$}, we consider \textit{precision} as the primary evaluation metric, \textit{recall} and \textit{F1 score} as complementary indicators, and the same calculation formula. We evaluate the ML-based judgments on identity pairs that are not used for model training. And the \textit{precision} is computed as the proportion of identity pairs identified as duplicates by the ML model that are confirmed as true duplicates according to the evaluation reference. We still adapt the McNemar's test, and conduct the same significance level for hypothesis testing. The null hypothesis \textit{(H$_{02}$)} states that there is no significant difference between ML-based approaches in duplicate developer identification, while the alternative hypothesis \textit{(H$_{12}$)} states that at least one ML-based approach differs significantly from others.

Moreover, we analyze the execution time and energy resource utilization of ML-based approaches (\textit{RQ$_{2.2}$}). We compare them with the cost of LLM-based approaches, and the goal is to explore whether ML-based approaches can provide a cost difference for duplicate developer identification, and determine their practical suitability for large-scale empirical mining studies \cite{dodge2022measuring}. We compute the costs for model training and inference. We aggregate execution time and system-level resource utilization metrics for each approach under the same computational environment.
To assess whether ML-based approaches outperform LLMs in terms of time and energy consumption, we compare the distributions of computational costs across the two categories. We conduct the paired t-test analysis~\cite{montgomery2019applied} to evaluate whether there is a significant difference in execution time and resource utilization. We perform hypothesis testing at a significance level (e.g., $\alpha = 0.05$). The null hypothesis \textit{(H$_{03}$)} states that there is no significant difference in time and energy consumption between ML-based approaches and LLMs for duplicate developer identification, while the alternative hypothesis \textit{(H$_{13}$)} states that ML-based approaches differ significantly from LLMs.

Additionally, we analyze how the precision of ML-based approaches changes with the amount of training data (\textit{RQ$_{2.3}$}). The objective of this analysis is to evaluate the data dependence of ML-based approaches on training data and determine whether increasing the volume of data can have a significant difference in precision.
For different predefined training data sizes, we compute the precision and compare the precision distributions. We conduct the hypothesis testing at a significance level ($\alpha = 0.05$). The null hypothesis \textit{(H$_{04}$)} states that there is no significant association between the amount of training data and the precision of ML-based duplicate developer identification, while the alternative hypothesis \textit{(H$_{14}$)} states that the precision of ML-based approaches differs with the amount of training data.

\textit{3) Cosine Similarity Augmentation (RQ$_3$):} We analyze whether adding cosine similarity based on developers' activity profiles could improve the quality of duplicate developer identification. We conduct this analysis on both LLM-based (\textit{RQ$_1$}) and ML-based (\textit{RQ$_2$}) identification. We set two conditions: (i) without cosine similarity (baseline setting) and (ii) with cosine similarity (augmented setting). We compute precision under these two conditions and compare its distributions to evaluate whether including cosine similarity affects precision.
To compare precision between the baseline and augmented settings, we use a paired t-test.
For \textit{RQ$_3$}, quality refers to the precision of ML-based duplicate developer identification.
To ensure comparability, we set only one variable, whether to increase cosine similarity, and keep all other conditions unchanged. We conduct the hypothesis testing at a significance level (e.g., $\alpha = 0.05$).
The null hypothesis \textit{(H$_{05}$)} states that there is no significant difference in the quality of duplicate developer identification when cosine similarity between developers is incorporated, compared to when it is not. The alternative hypothesis \textit{(H$_{15}$)} states that incorporating cosine similarity between developers has a significant difference in the quality of duplicate developer identification.

\section{Threats to Validity}
\label{sec:threats}

We discuss the threats to validity following the classification proposed by Wohlin et al.~\cite{wohlin2012experimentation}.

\textbf{Construct validity} In this work, developer identity is represented as a \textit{(name, email)} pair extracted from Git commit metadata. Such data cannot fully represent a natural person, as developers may change names, email addresses, or, conversely, developers with similar names can be encountered in a large project. Thus, the notion of a “duplicate developer identity” relies on human expert judgments used for precision estimation, which may be subjective in borderline cases (e.g., common names or ambiguous email patterns). To mitigate this threat, we rely on expert validation using a statistically grounded sampling strategy and focus on precision, which directly penalizes false positive merges that would introduce structural distortions in subsequent analyses. Moreover, we will leverage a conservative approach in which any uncertainty should result in pairs labeled as non-duplicate.

\textbf{Internal validity} A key threat arises from the blocking strategy used to reduce the number of candidate identity pairs, which may exclude true duplicate pairs that do not satisfy the initial string-based heuristics. While this could affect recall, our study explicitly prioritizes precision over completeness, as false positives are more harmful for organizational and architectural analyses.  ML-based approaches are trained on labels curated from LLM outputs, which introduces the risk of propagating systematic biases from LLMs into downstream models. Thus, our hypotheses do not deal with direct comparison of ML and LLM de-duplication performance, focusing instead on the time and effort required to obtain the results.. To reduce the risks, human experts are involved in validating sampled outputs, and all ML models are evaluated on held-out identity pairs not used during training.

\textbf{External validity} The primary study context is the Linux kernel, which is a large-scale, long-lived OSS project with a diverse and global contributor base. While this makes it a strong benchmark for scalability and heterogeneity, results obtained on this project may not directly generalize to smaller projects, projects with stricter contribution policies, or proprietary industrial settings. However, the focus of this work is on OSS mining scenarios where authoritative identity information is unavailable. If insufficient duplicate cases are observed in the Linux kernel, additional large OSS projects will be included to strengthen generalizability across different contributor ecosystems.

\textbf{Conclusion validity} Since this is a Stage 1 Registered Report we do not report threats to conclusion validity.

\section{Conclusion}
\label{sec:conclusion}
This study aims to address a key challenge in the Software Architecture (SA) mining field: accurately identifying duplicate developers in OSS projects. Since organizational and logical coupling metrics rely on developer-centric data, duplicate developer identities introduce systematic noise into collaboration networks and coupling metrics, potentially distorting architectural interpretations and downstream empirical conclusions. To this end, we propose LLM-based and classical ML-based approaches for duplicate developer identification, considering precision,  scalability, computational cost, and energy efficiency, and aim to construct a family of ML models for duplicate developer identification in OSS mining scenarios. The contribution of this work is to give a scalable developer identity de-duplication pipeline based on ML and LLM approaches, strengthening the developer-centric organizational and logical coupling analysis in the SA community.
\subparagraph*{Declaration on the use of generative AI}
We  used ChatGPT for suggestions on improving textual clarity. All research design, data analysis, interpretations, and manuscripts were created by the authors themselves.
\bibliographystyle{plainurl}
\bibliography{main}
\end{document}